\documentclass[final,3p,times,twocolumn]{elsarticle}
\usepackage{graphicx}
\usepackage{amsmath}
\usepackage{amssymb}
\usepackage{hyperref}

\begin{document}

\title{Structure, magnetic and dielectric properties in nano-crystalline Yb$_2$CoMnO$_6$}

\author{Ilyas Noor Bhatti}\address{Department of Physics, Jamia Millia Islamia University, New Delhi - 110025, India.}
\ead{inoorbhatti@gmail.com}
\author{Imtiaz Noor Bhatti\corref{cor2}}\address{Department of School Education, Government of Jammu and Kashmir, India.}
\ead{inbhatti07@gmail.com}
\author{Rabindra Nath Mahato}\address{School of Physical Sciences, Jawaharlal Nehru University, New Delhi - 110067, India.}
\author{M. A. H. Ahsan}\address{Department of Physics, Jamia Millia Islamia University, New Delhi - 110025, India.}

\begin{abstract}
Structural, magnetic and dielectric properties have been studied for Yb$_2$CoMnO$_6$.  Nano-crystalline sample of Yb$_2$CoMnO$_6$ synthesized by sol-gel method and structural analysis shows that the sample crystallizes in monoclinic crystal structure with \textit{P2$_1$/n} phase group. To understand the charge state of Co, Mn and Yb we have performed the XPS study. Magnetic study shows that the sample undergoes a paramagnetic to ferromagnetic phase transition around $T_c$ $\sim$56 K and an additional magnetic ordering at a lower temperature around 14 K due to ordering of Yb$^{3+}$ magnetic ions. Temperature dependent Raman study reveals that spin-phonon interaction is present in this material. Further, we have studied the dielectric properties of this material. We observed that the material shows a relaxation behavior that obeys the thermally activated relaxation mechanism.  Impedance spectroscopy reveals that the material shows non-Debye's behavior. AC conductivity study is performed to understand the conduction mechanics which involve the quantum mechanical tunneling phenomenon.
\end{abstract}

\maketitle
\section{Introduction} 

Materials which simultaneously shows ferroelectricity and magnetism are termed as multiferroic materials.\cite{kfwang, fiebig} Strong coupling of electric and magnetic properties can be utilized for vast industrial applications, for instance, storage devices, sensors, tunable microwave filters, etc.\cite{heron, seidel, hoff, sal} Recently, several double perovskite materials have exhibited coupled magnetic and electric phenomenon. Among them, a new double perovskite Ba$_2$FeMnO$_6$ have shown co-existence of electric and magnetic hysteresis and considered as a promising candidate for spintronic applications.\cite{ravi} The tunable magneto-electric effect has been observed in Y$_2$MnCrO$_6$ which have further intensified the quest of multiferroic materials in double perovskites.\cite{yong} Interestingly, novel high-temperature multiferroicity have been observed in a 3$d$-5$d$ based Bi$_2$(Ni/Mn)ReO$_6$ compound.\cite{lezai} Despite having an interesting physics and exotic phenomenon with promising potential for industrial applications these materials have not been well studied. 

Among the vast class of compounds 3$d$ based R$_2$CoMnO$_6$ (where R = rare earth elements) has received much attention of researchers due to their exotic properties such as E type ferromagnetic ordering, spin-phonon coupling, multiferroicity, magnetoelectricity, etc. In these double perovskites the spin magnetic moment of  Co$^{2+}$ and Mn$^{4+}$ interact via dominant super-exchange interaction and give rise to ferromagnetic ordering in these materials. However, in addition to this magnetic phase transition, the rare-earth ions also interact at low temperatures and in most cases align themselves in opposite direction to the Co/Mn sublattice and thus results in an antiferromagnetically ordered state. It is worth to mention some previous findings, for instance, Lu$_2$CoMnO$_6$ shows $E$-type magnetic ordering around 50 K with an anomaly in dielectric constant at same temperature which suggests some kind of magneto-electric interplay in this material.\cite{blasco} Further, this material shows pyroelectric properties and has shown negative magnetocapacitance.\cite{vilar} In another case, Er$_2$CoMnO$_6$ shows a ferromagnetic ordering of Co and Mn cations around 67 K with a low temperature ferrimagnetic ordering around 10 K activated by Er$^{3+}$ ions.\cite{banerjee} The pyroelectric and polarization properties of Y$_2$CoMnO$_6$ have been studied and confirms that no intrinsic magnetoelectric multiferroicity exists.\cite{blasco1} Magnetization and neutron study of single crystalline Yb$_2$CoMnO$_6$ shows $E$-type ferromagnetic ordering and also shows negative magnetocapacitance.\cite{blasco} Strong magnetic anisotropy and metamagnetic transition have been observed in a self flux-grown single crystal of Yb$_2$CoMnO$_6$. In this study, we aim to investigate the magnetic and dielectric properties of nano-crystalline Yb$_2$CoMnO$_6$ and compare the results with bulk study to identify the effect of reduced dimensions.

In the present study, we report structural, magnetic, dielectric and transport properties of nano-crystalline Yb$_2$CoMnO$_6$. The structural analysis shows the sample is in single phase and adopts the monoclinic crystal structure with \textit{P2$_1$/n} space group. Magnetization study reveals that the Yb$_2$CoMnO$_6$ is a ferromagnetic material that undergoes PM-FM phase transition around $T_c$ $\sim$56 K. The material also shows antiferromagnetic ordering at low temperature. Raman study shows spin-phonon coupling is present in this material. Dielectric measurement shows a strong dispersion in dielectric constant and tangent loss shows a relaxation phenomenon. The impedance spectroscopy shows that Yb$_2$CoMnO$_6$ does not follow Debye's model. The Nyquist plot analysis shows non-Debye's behavior Yb$_2$CoMnO$_6$. AC conductivity shows strong frequency dependency at the higher frequency limit. The conductivity analysis shows that the conduction mechanism involves the quantum mechanical tunneling phenomenon.
 
\section{Experimental details}
Sol-gel method has been employed to synthesize the nano-crystalline Yb$_2$CoMnO$_6$. Starting ingredients with high purity (99.9$\%$) from Alpha Aesar were used. First, we have prepared solutions of each compound in separate beakers. We have dissolved Mn(CH$_2$CO$_2$).4H$_2$O, Co(NO$_3$)$_2$.6H$_2$O and nitric acid in water with continuous stirring till the solution becomes clear. However, Yb$_2$O$_3$ is insoluble in water to make a clear solution we have added nitric acid drop by drop to the beaker containing Yb$_2$O$_3$ and water with continuous stirring at 95 $^o$C. After continuous staring for 20 minutes Yb$_2$O$_3$ dissolves completely in dilute nitric acid and the solution becomes clear. Then all the solutions were poured into the 500 ml beaker and filled to 400 ml with distilled water. This final solution is then kept on magnetic starrier at 90 $^o$C for 24 hours for suitable reaction time. The solution then turned into a gel and finally heated up to form a foam. The foam is collected from the beaked grounded well and heated at 600 $^o$C. The collected powder is then grounded again for 30 minutes and sintered at 900 $^o$C for 12 hours. The obtained powder is collected and characterized by X-ray diffraction using Rigaku mini flex 600 diffractometer. X-ray photoelectron spectroscopy (XPS) is performed to understand the charge state of cations. The XPS measurements were performed with base pressure in the range of $10^{-10}$ mbar using a commercial electron energy analyzer (Omnicron nanotechnology) and a non-monochromatic Al$_{K\alpha}$ X-ray source (h$\nu$ = 1486.6 eV). The XPSpeakfit software has been used to analyze the XPS data. The samples used for XPS study are in pellet form where an ion beam sputtering has been done on the samples to expose clean surface before measurements. The magnetic measurements were done on powder using a vibrating sample magnetometer by Cryogen.Inc. Temperature dependent Raman spectra have been collected using Diode based laser ($\lambda$ = 473 nm) coupled with a Labram-HR800 micro-Raman spectrometer. It is a single spectrometer with 1800 groves/mm grating and a peltier cooled CCD detector with an overall spectral resolution of $\sim$1 cm$^{-1}$. For the low temperature Raman measurements, the material has been mounted on a THMS600 stage from Linkam UK, with temperature stability of $\pm$0.1 K. Dielectric measurements in the frequency range from 1 Hz to 5.5 MHz were performed using a computer controlled dielectric set up on a closed-cycle refrigerator with operating temperature range of 20 to 300 K.

\begin{figure}[t]
    \centering
        \includegraphics[width=8cm,height=10cm]{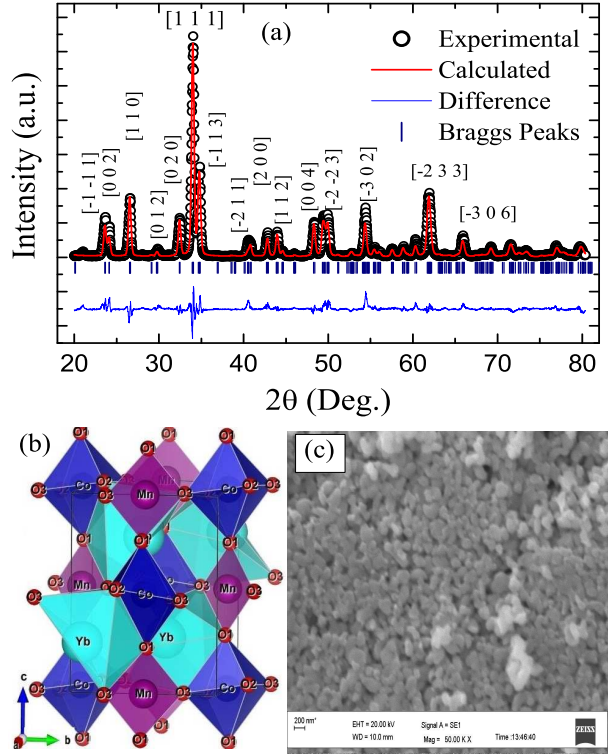}
\caption{(Color online) (a) X-ray diffraction pattern along with Rietveld refinement for nano-crystalline Yb$_2$CoMnO$_6$, peak indexing is represented for main peaks. (b) Shows the monoclinic unit cell structure obtained from refinement data. (c) Representative SEM image use to estimate average partical size. }
    \label{fig:Fig1}
\end{figure}

\section{Result and Discussions}
\subsection{Structural study}
Fig. 1a shows the X-ray diffraction pattern along with Rietveld refinement for nano-crystalline Yb$_2$CoMnO$_6$ double perovskite. In this figure the black open circles are experimental data, the bold red solid line is the calculated model, the blue weak line is the difference in experimental and calculated pattern. The fitting parameter \textit{goodness of fit} $\chi^2$ = 1.93 which is quite reasonable. Further, from fitting we obtained R$_{wp}$/R$_{exp}$  ratio  $\sim$ 1.38 which is reasonably good and acceptable.\cite{ilyas1, bhatti1, bhatti2} These parameters show the Rietveld refinement is reasonably good. Braggs peaks are also shown with navy vertical bars. The Rietveld analysis on XRD data has shown that the sample is chemically pure and in single phase. Peak indexing is given in Fig. 1a for main peaks. The sample crystallized in monoclinic crystal structure with  $P2_1/n$ space group. The lattice parameters are $a$ = 5.1535(8) $\AA$ $b$ = 5.5215(7) $\AA$ $c$ = 7.3832(5) $\AA$ and $\beta$ = 90.29(1) $^o$ with the unit cell volume 210.08(9) $\AA^3$. The crystal structure unit cell is shown in Fig. 1b for this sample, it is evident from the figure that the Co and Mn atoms are alternatively placed. Further, to measure the average particle size we used the scanning electron microscope to obtain the micrograph of nano-particles as shown in Fig. 1c. The imageJ software is used to analyze the SEM micrograph  and we found that the nano-particle of Yb$_2$CoMnO$_6$ are obtained with average particle size is $\sim$ 70 nm.
\begin{figure}[t]
    \centering
        \includegraphics[width=7cm,height=13cm]{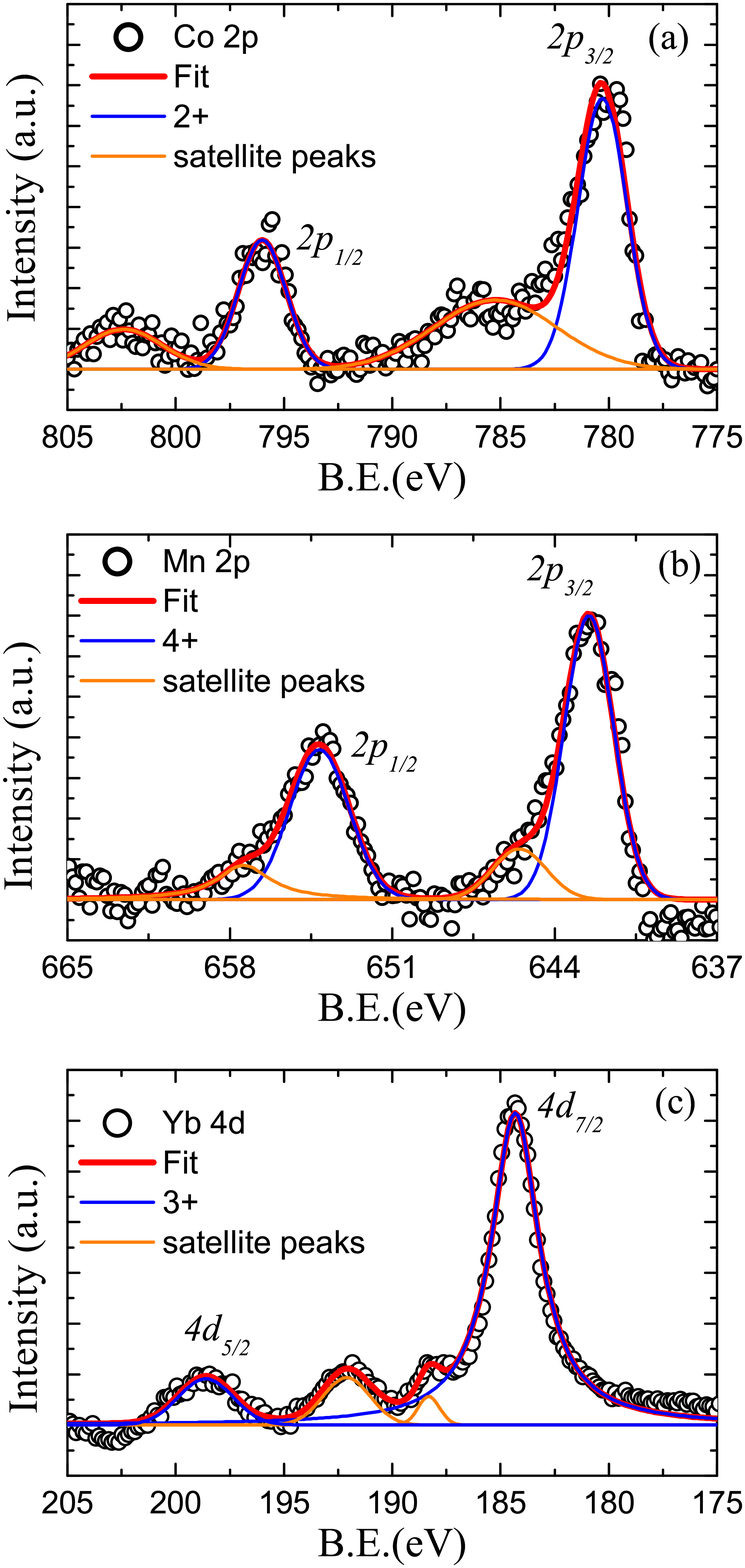}
\caption{(Color online) (a) show the XPS core level spectra of Co 2$p$ (b) show the XPS core level spectra of Mn 2$p$. (c) shows the XPS core level spectra of Yb 4$d$ for Yb$_2$CoMnO$_6$. In the figure the red solid line is the overall envelope of the XPS spectrum and the other colored solid lines are the other respective fitted peaks.}
    \label{fig:Fig2}
\end{figure}

\subsection{X-ray photoelectron spectroscopy (XPS)}
The physical properties of a compound are largely described by the oxidation state of cations present in the material. XPS is a vital tool to understand the charge state of cations. We have employed the XPS to study the cationic charge state of Co, Mn and Yb in Yb$_2$CoMnO$_6$. XPS spectrum of Co 2$p$ is shown in Fig. 2a in which the open black circles are the experimental data the red line is the overall envelope of the spectrum the solid blue lines are the Co 2$p$ peaks whereas orange solid lines are the satellite peaks fitted using XPS PEAKFIT 4.1. It is evident from Fig. 2a there are two peaks located at 780.7 eV and 796.01 eV for CO 2$p$$_{3/2}$ and Co 2$p$$_{1/2}$  respectively resulting from the spin-orbital splitting of 2$p$ orbital with 15.3 eV. Besides the Co 2$p$ there are two other peaks marked as satellite positioned at 787 eV and 802.5 eV. The results are consistent with literature.\cite{wang, qiu, xia}  Beside the Co 2$p$ peaks two satellite peaks have also been observed close to Co 2$p$ peaks. The peaks locations of Co 2$p$ core level indicate that the Co cations are present in +2 oxidation states.

The measured XPS spectrum for Mn 2$p$ core levels along with peak fitting is shown in Fig. 2b, where the open black circles are the experimental data the red line is the overall envelope of the spectrum the solid blue lines are the Mn 2$p$ peaks are shown. The Mn 2$p$ spectrum shows two distinct peaks located at 642 eV and 654 eV corresponding to Mn 2$p$$_{3/2}$ and 2$p$$_{1/2}$ resulting from the spin-orbital splitting of 2$p$ orbitals with splitting energy of 12 eV. The peak positions and splitting energy is reveals that the Mn cation is present in +4 oxidation state and is in agreement with literature.\cite{ida, sachoo, cao}

XPS spectra of Yb 4$d$ core level along with the fitting of peaks are shown in Fig. 2c where the open black circles are experimental data, the solid red line is the overall envelope of the XPS spectrum and solid blue lines are the Sm 3$d$ orbitals. It is evident from the figure that the two distinct spin-orbital split peaks Yb-4d$_{5/2}$ and Yb-4d$_{3/2}$ peaks are located at 186 eV and 199.6 eV, respectively with at spin-orbital splitting energy of 13.6 eV. \cite{sarma, tiana} However, besides these two peaks there are more small peaks attributed to the presence of Yb$^{2+}$ oxidation state in the material. The detailed analysis of XPS spectrum reveals that the majority of Yb cations are present in +3 oxidation states whereas small Yb$^{2+}$ cations are also present.\cite{tiana} It is conclusive from the XPS study that Co, Mn and Yb cations are in 2+, 4+ and 3+ oxidation states respectively.

\begin{figure}[t]
    \centering
        \includegraphics[width=8cm]{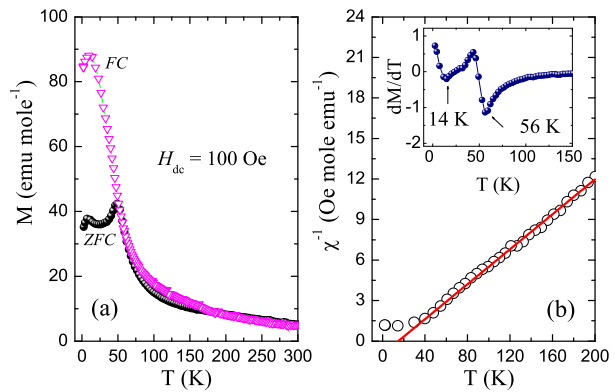}
\caption{(Color online) (a) Temperature dependent magnetization data $M(T)$ shown for Yb$_2$CoMnO$_6$ measured at different fields. (b) $M(T)$ data plotted in terms of inverse susceptibility $\chi^{-1}$, solid line is fitting due to Curie Weiss Law. Inset shows $dM/dT$ vs $T$ plot showing $T_c$ for Yb$_2$CoMnO$_6$.}
    \label{fig:Fig3}
\end{figure}

\subsection{Magnetization study}
Temperature dependent magnetization ($M(T)$) measurements were performed both in the zero field cooled ($ZFC$) and field cooled ($FC$) mode as shown in Fig. 3a. $ZFC$ and $FC$ data is recorded in a temperature range of 300 K to 2 K under an applied dc magnetic field ($H_{dc}$) of 100 Oe. The figure shows that with the decrease in temperature the magnetic moment ($M$) in $M(T)$ curve remains steady initially. However, around 60 K the magnetic moment in both the ($FC$) and ($ZFC$) curve began to increase with a further decrease in temperature. It is expected that the rise in magnetic moment is due to establishment of magnetic exchange between Co$^{2+}$-Mn$^{4+}$ in Yb$_2$CoMnO$_6$ double perovskite which is superexchange in nature and give rise to ferromagnetic ordering below 56 K. This sharp rise in the magnetic moment below 56 K is marked by paramagnetic ($PM$) to ferromagnetic ($FM$) phase transition in Yb$_2$CoMnO$_6$. It is seen in Fig. 3a that the large bifurcation appears in $M_{ZFC}$ and $M_{FC}$ with further decreasing temperature. A peak like behavior around $T_c$ is observed in $M_{ZFC}$ curve and with further decreasing temperature the $M_{ZFC}$ began to decrease, such type of behavior is seen in many perovskite compounds.\cite{bhatti1, bhatti2, ilyas2, ilyas3, renu, bhatti3, ilyas4}  Whereas $M_{FC}$ continuously increases with decreasing temperature. However, at low temperature below $\sim$15 K there appears a peak in $M_{FC}$ and $M_{ZFC}$ which is related to the ordering of Yb$^{3+}$ ions at low temperature. The bifurcation of $ZFC$ and $FC$  in oxides with composition La$_{0.7-x}$Dy$_{x}$Ca$_{0.3}$Mn(Fe)O$_3$ has been studied in detail.\cite{bhar1,bhar2} The observation of such bifurcation is linked with the spin glass phase in this compound. However, the observed bifurcation is influenced by composition as reported and suppression of spin glass phase is observed. However, to better understand the bifurcation in $ZFC$ and $FC$ of Yb$_2$CoMnO$_6$ dynamic susceptibility measurement is required, where frequency dependent ac susceptibility can shed light on the presence of glassy phase in this material. To estimate the transition temperature across which magnetic phase transition from $PM-FM$ appears we have plotted the $dM/dT$ vs $T$ where the point of inflection in this plot gives the $T_c$. Inset Fig. 3b shows two points of inflection, one around 56 K the $T_c$ corresponding to the ferromagnetic ordering of Co$^{2+}$-Mn$^{4+}$ whereas the other appears at 14 K which is attributed to ordering of Yb$^{3+}$ with net moment opposite to Co/Mn sublattice.

To further understand the magnetic behavior, we have plotted the temperature dependent magnetization data in terms of inverse magnetic susceptibility i.e. $\chi^{^{-1}}$ vs $T$ as shown in the main panel of Fig. 3b.  In the paramagnetic region above $T_c$, the inverse magnetic susceptibility can be fitted with conventional Curie Weiss law described as:\cite{chikazumi, cullity}

\begin{eqnarray}
\chi = \frac{C}{T - \theta_P}
\end{eqnarray}

Here $C$ is Curie Constant and $\theta_P$ paramagnetic Curie temperature. We have observed that the inverse susceptibility is varying linearly with the temperature above $T_c$. The susceptibility data of Yb$_2$CoMnO$_6$ in the temperature range 70 K to 200 K is fitted with Curie Weiss law given in Eq. 1. The fitting parameters obtained from fitting of $\chi^{-1}$ with Eq. 1 in Fig. 3b were used to calculate Curie constant and $\theta_P$. The obtained values of C and  $\theta_P$ for Yb$_2$CoMnO$_6$ are 15.470(2) emu K mole$^{-1}$ Oe$^{-1}$ and 14.08 K respectively. The positive value of $\theta_P$ signifies that the ferromagnetic ordering is present in Yb$_2$CoMnO$_6$ sample. Further, the effective magnetic moment is calculated using formula  $\mu_{eff}$ = $\sqrt{3 C k_{B}/N}$ where $C$ is obtained from the slope of Curie Weiss fitting in Fig. 3b. The value of $\mu_{eff}$ is 11.12(6) $\mu_B$/f.u.

\begin{figure}[t]
    \centering
        \includegraphics[width=8cm]{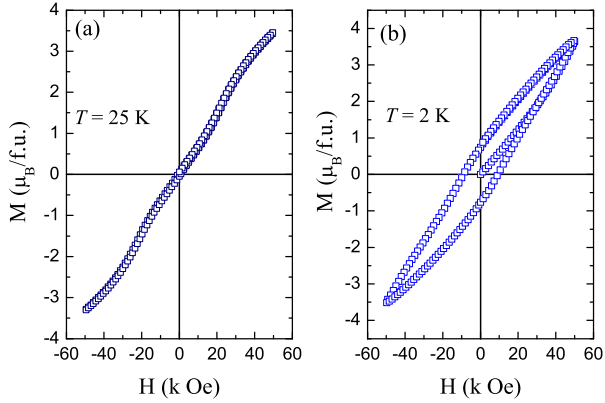}
\caption{(Color online) Isothermal magnetization data collected at 25 K (see (a)) and 2 K (see (b)) in an applied magnetic field of up to $\pm$ 50 kOe for Yb$_2$CoMnO$_6$.}
    \label{fig:Fig4}
\end{figure}

\begin{figure*}[th]
    \centering
        \includegraphics[width=14cm]{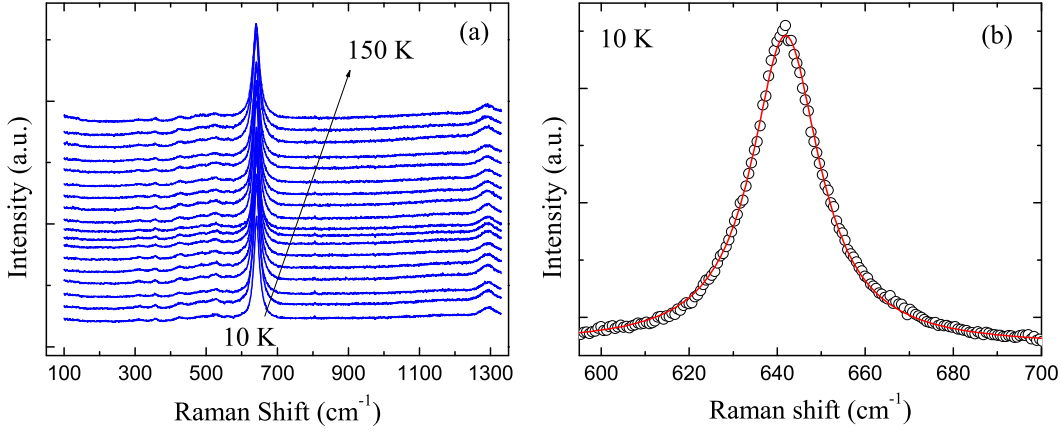}
\caption{(Color online) (a) Raman spectra of Yb$_2$CoMnO$_6$ measured at different temperatures.   (b) shows the line shape and its Lorentzian fitting of B$_{2g}$ Raman mode at 641 cm$^{-1}$.}
    \label{fig:Fig5}
\end{figure*}

Isothermal magnetization i.e. the variation of magnetization against temperature $M(H)$ data have been collected at 25 K and 2 K up to $\pm$50 kOe applied magnetic field as shown in Fig. 4a and 4b respectively. $M(H)$ curve does not show any hysteresis at 25 K, however, we observe a slight slope change around 20 kOe in both positive and negative magnetic field directions. The $M(H)$ at 2 K is completely different from $M(H)$ curve at 25 K. $M(H)$ at 2 K shows a large hysteresis which suggests the overall ferromagnetic nature of Yb$_2$CoMnO$_6$. Further, the magnetic moment does not show any signs of saturation even at the highest applied magnetic field of 50 kOe at both 2 K and 25 K. From the $M(H)$ curve at 2 K we found that the magnetic moment at 50 kOe is about 3.5 $\mu$$_B$/f.u. whereas the remanent magnetization and coercive force are 0.8387 $\mu$$_B$/f.u. and 9065 Oe respectively.

\subsection{Temperature dependent Raman study}
Temperature dependent Raman spectra taken at selective temperatures across magnetic transition is shown in Fig. 5a. The Raman spectra are taken in the temperature range from 10 K to 150 K across the magnetic transition whereas the Raman spectra are taken in the close temperature intervals across $T_c$. It is evident from Fig. 5a the important feature in the Raman spectra are the prominent Raman modes at 641 and 496 cm$^{-1}$ corresponding to B$_{2g}$ stretching mode and A$_{1g}$ breathing mode respectively. These Raman modes are due to stretching, bending and rotation of (Co/Mn)O$_6$ octahedra. It is confirmed from the theoretical lattice dynamics that the strong sharp peak around 636 cm$^{-1}$ originates from symmetric stretching of the (Co/Mn)O$_6$ octahedra, whereas the band at around 496 cm$^{-1}$ describes a mixed type vibration of antisymmetric stretching and bending.\cite{ilive} Additionally, the modes at $\sim$1278 cm$^{-1}$ represent the second-order overtones of the breathing mode.\cite{meyer} The temperature dependent Raman spectra show variation in the peak positions and line width corresponding to each mode. The line shape along with the Lorentzian fitting is shown for phonon mode at 641 cm$^{-1}$ is shown in Fig. 5b.

\begin{figure}[th]
    \centering
        \includegraphics[width=8cm]{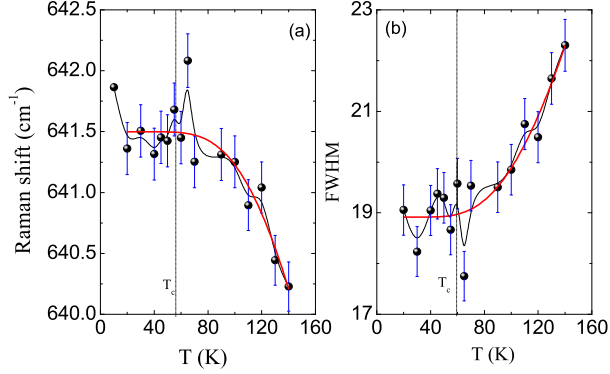}
\caption{(Color online) Temperature Variation of (a) Raman shift (b) FWHM for Raman mode at 641 cm$^{-1}$ corresponding to stretching of Co/MnO$_6$ for Yb$_2$CoMnO$_6$. The solid line is fitting due to Eq. 2}
    \label{fig:Fig6}
\end{figure}

To understand the presence of spin-phonon coupling in Yb$_2$CoMnO$_6$, we have analyzed the Raman data of present series with the following anharmonic decay model:\cite{ilyas3, granado}

\begin{eqnarray}
\omega(T)= \omega{_0} - C\left[1 + \frac{2}{exp\left(\frac{\hbar\omega_0}{2k_BT})\right) - 1}\right]
\end{eqnarray}

where $\omega_{0}$ and C are the intrinsic frequency of the optical mode and anharmonic coefficient, respectively. $\omega(T)$ describes the expected temperature dependence of a phonon mode frequency due to anharmonic phonon-phonon scattering. 

The temperature dependent Raman shift ($\omega$(T)) at 641 cm$^{-1}$ is shown in Fig. 6a. We have fitted the Raman mode with anharmonic decay model shown in Eq.2. The $\omega$(T) fitted well with this model above $T_c$, however, we have observed that the $\omega$(T) deviates from anharmonic behavior around $T_c$ shown in the figure where $T_c$ guided by the dotted line. The deviation of $\omega$(T) around $T_c$ appears due to phonon renormalization caused by ferromagnetic ordering. This result suggests a spin-phonon coupling in  Yb$_2$CoMnO$_6$. Such behavior has also been reported for many other compounds.\cite{meyer, granado, laver} It is further noted that the temperature dependent line width of this phonon mode decreases with decreasing temperature as shown in Fig. 6b. It is evident from the figure that the line width also shows a deviation from anharmonic behavior across $T_c$. This deviation is attributed to the additional scattering mechanism involved such as spin-phonon coupling. Thus it is evident from Raman study that the spin-phonon coupling is present in Yb$_2$CoMnO$_6$.

\begin{figure*}[th]
    \centering
        \includegraphics[width=14cm]{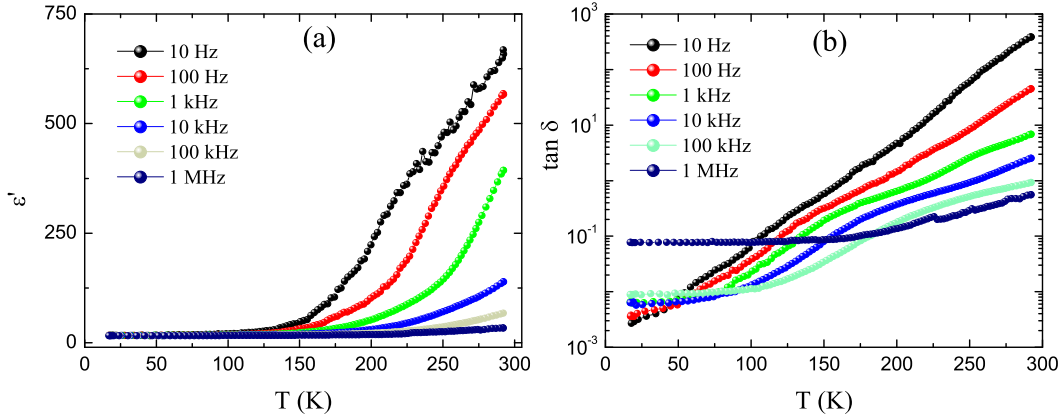}
\caption{(Color online) Temperature dependent (a) real part of complex dielectric permittivity ($\epsilon$$^\prime$) (b) loss tangent (tan $\delta$) measure for Yb$_2$CoMnO$_6$ in the temperature range of 20 K to 300 K at various frequencies.}
    \label{fig:Fig7}
\end{figure*}

\begin{figure}[t]
    \centering
        \includegraphics[width=8cm]{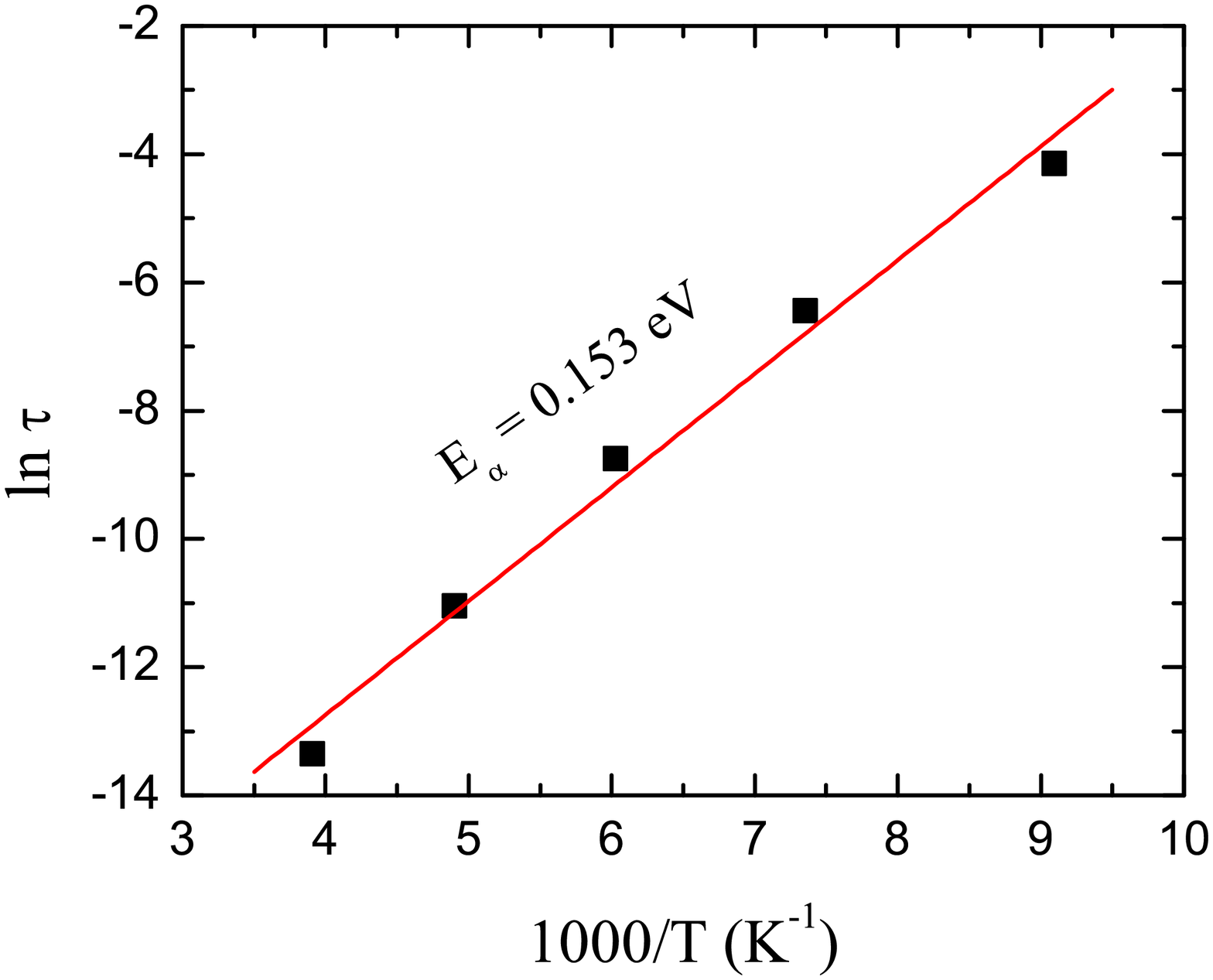}
\caption{(Color online) Variation of relaxation time against normalized temperature i.e ln $\tau$ vs 1000/T obtained from tan $\delta$ plot.}
    \label{fig:Fig8}
\end{figure}

\begin{figure*}[th]
    \centering
        \includegraphics[width=14cm]{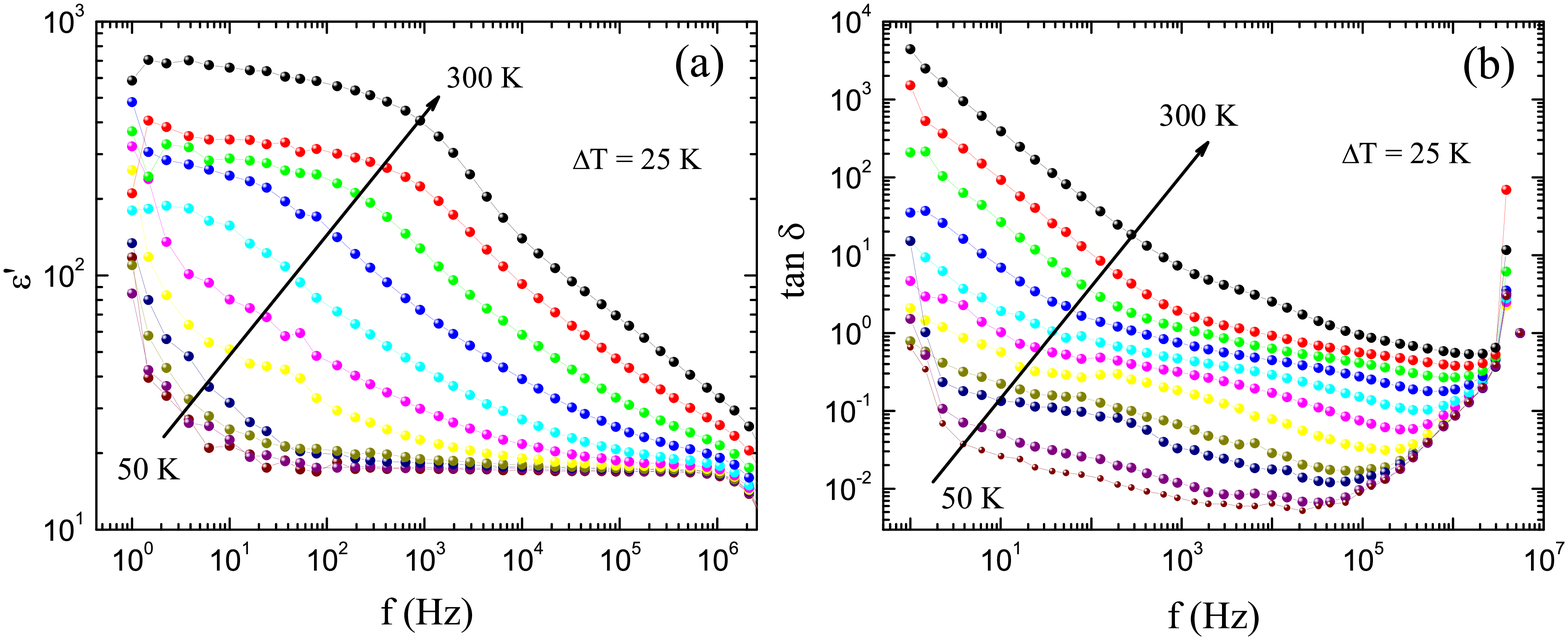}
\caption{(Color online) Frequency dependent (a) real part of complex dielectric permittivity ($\epsilon$$^\prime$) (b) loss tangent (tan $\delta$) measure for Yb$_2$CoMnO$_6$ at various temperatures between 25 K and 300 K in the frequency range 1 Hz to 5.5 MHz.}
    \label{fig:Fig9}
\end{figure*}

\subsection{Dielectric study}
Fig. 7a and 7b represent the temperature dependent real and imaginary part of complex dielectric permittivity $\epsilon^{\prime}$ and tan $\delta$  respectively measured in the temperature range 20 K to 300 K at different frequencies for Yb$_2$CoMnO$_6$. Further, for relaxor systems with relaxation mechanisms, each relaxation component will correspond to plateaus in $\epsilon^{\prime}$(T) and respond with peaks in tan $\delta$. For this material, we have observed that with increasing temperature the $\epsilon^{\prime}$ at low temperatures  $\epsilon^{\prime}$ increases slowly however with increasing temperature $\epsilon^{\prime}$ increases sharply. Further, with increasing frequencies, $\epsilon^{\prime}$ decreases sharply. The higher value of $\epsilon^{\prime}$ at low frequency is attributed to the accumulation of the charges at grain boundaries  On careful observation of tangent loss curve tan $\delta$, it is seen that there is a broad hump at low temperature which is a feature of relaxor phenomenon. The observed relaxation is frequency dependent and shifts to higher temperatures with increasing frequency.

The relaxation mechanisms and its origin can be analyzed by fitting the peaks in tan $\delta$ with the Arrhenius law given by, $\tau_{tan  \delta}$ = $\tau_0$ exp(-$E_a$/$K_B$$T$) where, $T$ is the temperature where peak occurs in tangent loss curve at a particular frequency $\tau_{tan  \delta}$, $\tau_0$ and $E_a$ are characteristic relaxation time and activation energy respectively and, $k_B$ is the Boltzmann constant.\cite{sch, kao, mansuri, elli, bhatn}

Fig. 8 shows the variation of relaxation time with absolute temperature i.e. ln$\tau$ vs 10$^{3}$/T. It is evident from the figure that the relaxation time is well fitted with Arrhenius law which suggests the thermally activated relaxation mechanism. From the fitting parameters, we have calculated activation energy E$_a$ = 0.153 eV.
To further understand the dielectric response we have measured the frequency dependent $\epsilon$$\prime$ and tan $\delta$ over the frequency range 1 Hz to 5.5 MHz for Yb$_2$CoMnO$_6$ at different temperatures. In Fig. 9a we observed that Yb$_2$CoMnO$_6$ exhibits high dielectric constant at low frequency and high temperature. The dielectric spectrum is shown in Fig. 9a clearly shows two plateaus well separated by dispersion indicated by the arrow (Fig. 9a). The separate plateau in Fig. 9a are attributed to static and optical dielectric constant. Further, the frequency dependent dielectric constant shows the dispersion which moves to the higher frequency with increasing temperature. Fig. 9b shows the dielectric loss as a function of frequency measured at selective temperatures. The dielectric loss is relevant to the permittivity of the material.

\subsection{Impedance spectroscopy}
Fig. 10a shows the real part of complex impedance ($Z^{\prime}$) plotted as a function of frequency in the frequency range 1 Hz to 5.5 MHz at various temperatures between 50 K to 300 K. For clarity both the axis are in logarithmic scales. It is quite evident from the figure that the $Z^{\prime}$ decreases with increasing temperature. At low temperature $Z^{\prime}$ gradually decreases with increasing frequency, however, at the temperature above 100 K the $Z^{\prime}$ initially remains independent of frequency then at the higher frequency it began to decreases. Further, the frequency independent region moves to the high frequency region with increasing temperature. Further, it is observed that at high frequency and high temperature the $Z^{\prime}$ almost similar. This feature possibly is due to the release of accumulated space charges at high temperatures hence contribute to the enhancement of conduction in this material at high temperatures. The imaginary part of impedance ($Z^{\prime\prime}$) is shown in Fig. 10b for a wide frequency range. $Z^{\prime\prime}$ shows a peak which attains $Z^{\prime\prime}_{max}$ for all the curves measured at different temperatures, further it is evident that the peak moves towards higher frequency with increasing temperature.

\begin{figure}[th]
    \centering
        \includegraphics[width=8cm]{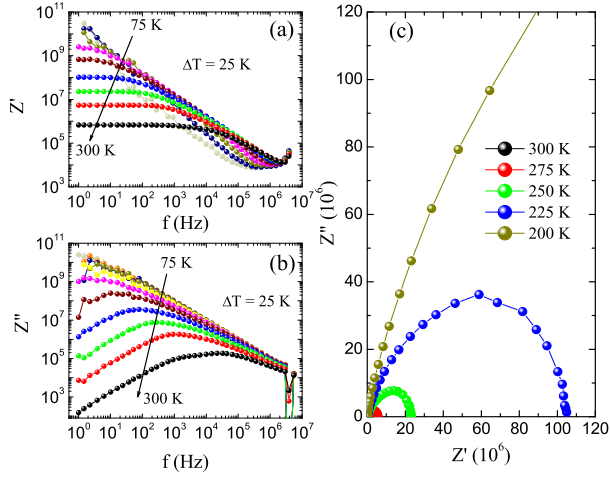}
\caption{(Color online) (a) Frequency dependent real part of impedance $Z^{\prime}$ measure at different temperatures. (b) Frequency dependent imaginary part of impedance $Z^{\prime\prime}$ measure at various temperatures. (c) real $Z^{\prime}$ and imaginary part $Z^{\prime\prime}$ plotted in terms of Nyquist plot $Z^{\prime}$ vs $Z^{\prime\prime}$.}
    \label{fig:Fig10}
\end{figure}

\begin{figure}[th]
    \centering
        \includegraphics[width=8cm]{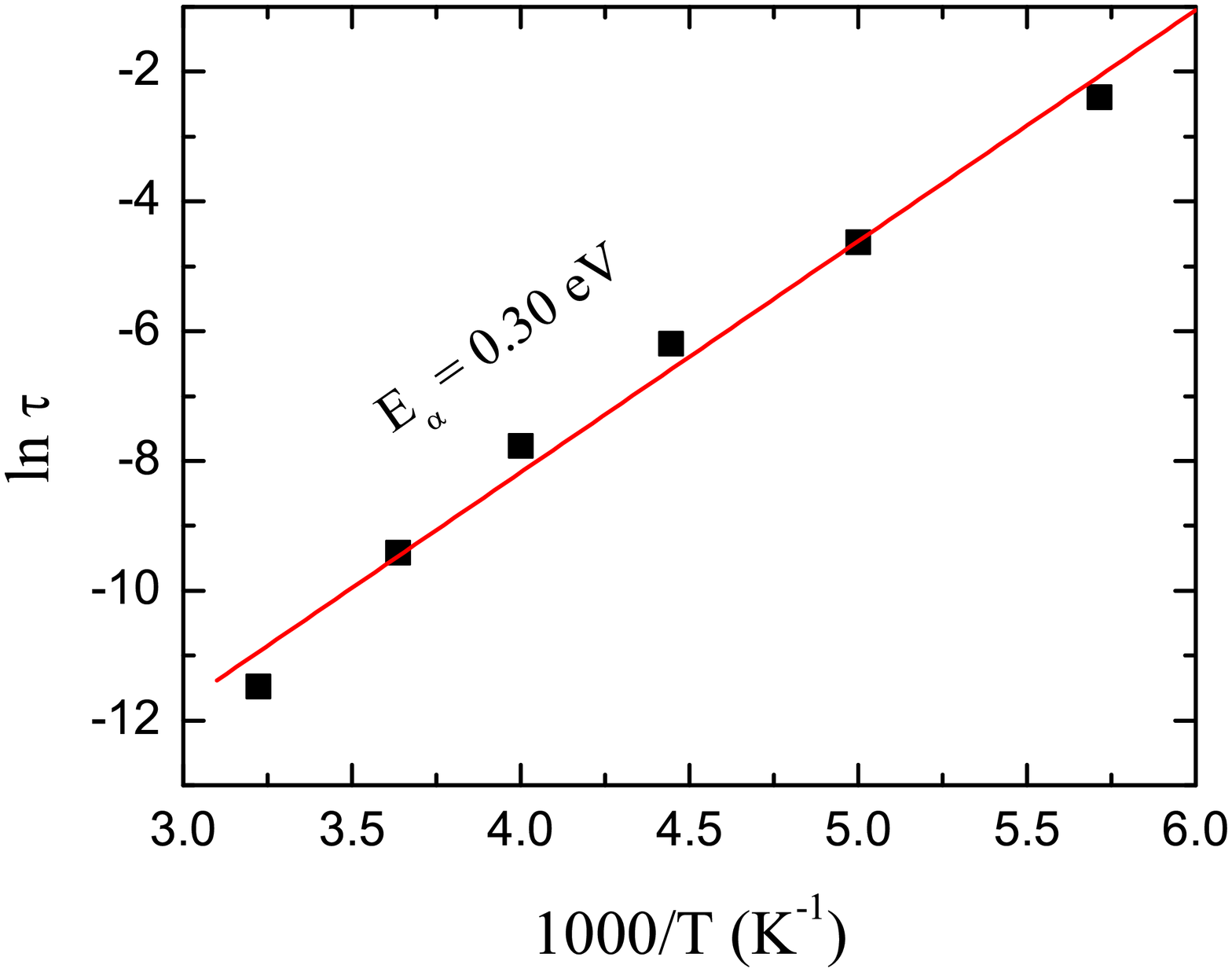}
\caption{(Color online) Variation of relaxation time against normalized temperature i.e ln $\tau$ vs 1000/T obtained from Z$^{\prime\prime}$ plot.}
    \label{fig:Fig11}
\end{figure}

We know that the most probable relaxation time ($\tau$) can be determined for relaxation system by identifying the position of the loss peak in the $Z^{\prime\prime}$ vs log ($f$) plots using the relation:\cite{sch,kao, kim,  elli, bhatn}

\begin{eqnarray}
\tau = \frac{1}{\omega} = \frac{1}{2 \pi f}
\end{eqnarray}

where $\tau$ is relaxation time and $f$ is the relaxation frequency. 
To further understand the relaxation behavior we have plotted the relaxation time $\tau$ vs inverse temperature 10$^3$/$T$ (K$^{-1}$). Fig. 11 shows the temperature variation of $\tau$, it is observed that the relaxation time follows Arrhenius behavior given as: \cite{elli, bhatn, ali}
 
\begin{eqnarray}
\tau_b = \tau_0 exp \left( \frac{-E_a}{k_BT} \right)
\end{eqnarray}

where $\tau_0$ is the pre-exponential factor, k$_B$ the Boltzmann constant and $T$ the absolute temperature. From the fitting parameters, the activation energy (E$_a$) has been calculated and is found to be 0.15 eV.
\begin{figure*}[th]
    \centering
        \includegraphics[width=12cm]{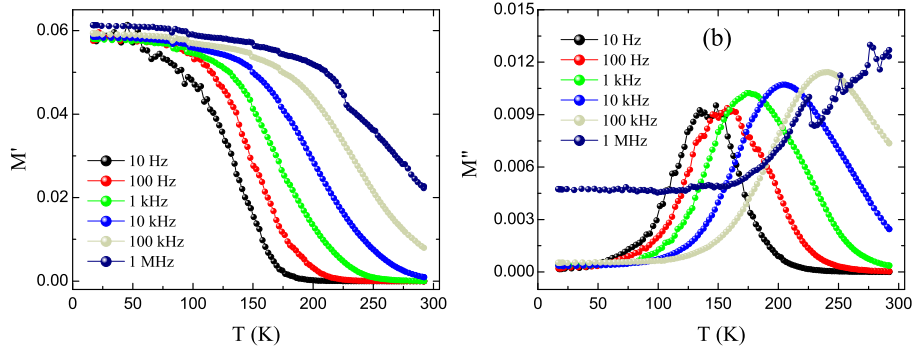}
\caption{(Color online) (a) Variation of real part of electrical modulus ($M^\prime$) with temperature. (b) Imaginary part of electrical modulus $M^{\prime\prime}$ as a function of temperature.}
    \label{fig:Fig12}
\end{figure*}

\begin{figure}[th]
    \centering
        \includegraphics[width=8cm]{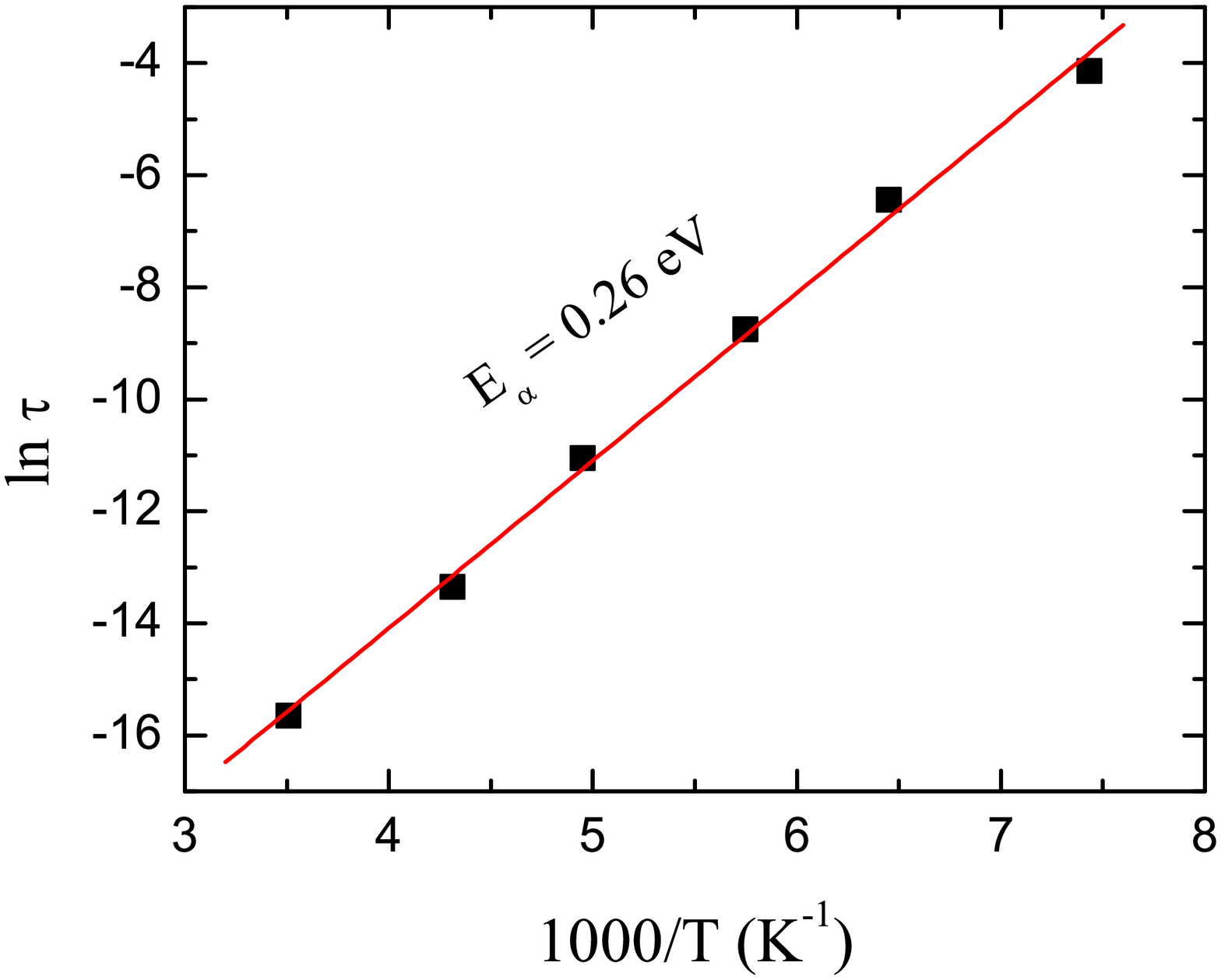}
\caption{(Color online) Variation of relaxation time against normalized temperature i.e ln $\tau$ vs 1000/T obtained from M$^\prime\prime$ plot.}
    \label{fig:Fig13}
\end{figure}
Fig. 10c shows the $Z^{\prime}$ vs $Z^{\prime\prime}$ in the form of Nyquist plots at some selective temperatures measure in the wide frequency range 1 Hz to 5.5 MHz. It is quite evident from the figures that the plot gives the semicircle in the whole range of temperature. The experimentally obtained impedance data for Yb$_2$CoMn$_6$ is plotted on the complex plane in terms of the Nyquist plot i.e. Z$^\prime$ vs Z$^{\prime\prime}$.  The semicircle in the Nyquist plots has depressed which suggests that the non-Debye type of relaxation is found in this material. It also manifests that there is a distribution of relaxation time instead of a single relaxation time in the material. Further, it is observed that the radius of semicircles decreases with increasing temperature.

\subsection{Electrical modulus}
Information of interface polarization, relaxation time, electrical conductivity and grain boundary
conduction effects, etc can be deduced from the electrical modulus of materials. Figs. 12a and 12b show the temperature dependent real ($M^{\prime}$) and imaginary ($M^{\prime\prime}$) part electrical modulus obtained at selective frequencies for Yb$_2$CoMnO$_6$ in the temperature range of 20 K to 300 K. 

Fig. 12b shows variation of the $M^{\prime\prime}$ with frequency at selected temperatures. Once again, $M^{\prime\prime}$ spectroscopy plot reveals relaxation phenomena in the material. The maximum value ($M^{\prime\prime}$) in the $M^{\prime\prime}$ vs log$f$ plot shows peak shifts to the higher frequency, which suggests that hopping of charge carriers is predominantly thermally activated. The asymmetric broadening of the peak indicates the spread of relaxation with different time constants, which once again suggests the material is non-Debye-type.

To understand the relaxation mechanism we can have plotted the relaxation time as a function of absolute temperature i.e. $\tau$ vs 10$^{3}$/T shown in Fig. 13. The relaxation time is distributed and suggests that the relaxation mechanism is thermally activated in nature and follows Arrhenius behavior given by:\cite{elli, bhatn, ali}

\begin{eqnarray}
\tau  =   \tau_{0}exp\left(\frac{-E_a}{k_BT}\right)
\end{eqnarray}
where $\tau_{0}$ is pre-exponent factor, $k_B$ is Boltzmann constant and $E_a$ is the activation energy.
we have fitted the data with above Eq. as shoown in Fig. 13 and found that the data fitted well. From the fittng parameter we have calculated the activation energy E$_a$ = 0. 26 eV

\begin{figure}[th]
    \centering
        \includegraphics[width=8cm]{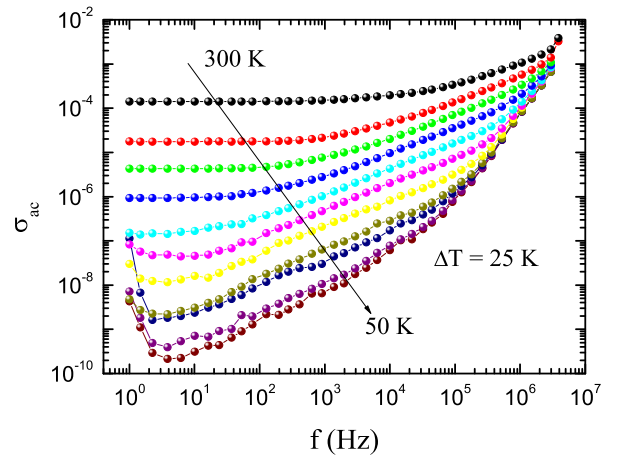}
\caption{(Color online) Frequency dependence plot of the ac conductivity ($\sigma_{ac}$) for temperatures ranging from 50 K to 300 K are shown for Yb$_2$CoMnO$_6$.}
    \label{fig:Fig14}
\end{figure}

\begin{figure*}[th]
    \centering
        \includegraphics[width=12cm]{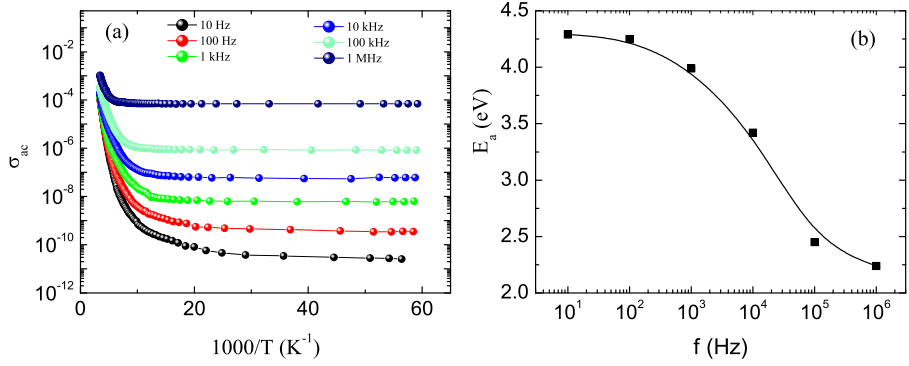}
\caption{(Color online) (a) The variation of $\sigma_{ac}$ with absolute temperature (10$^3$/T) is shown for Yb$_2$CoMnO$_6$.The solid lines are due to fitting wth Eqn. 7. (b) Variation of activation energy as a function of frequency for Yb$_2$CoMnO$_6$..}
    \label{fig:Fig15}
\end{figure*}

\subsection{Electric conductivity}
To further understand the charge hopping and electrical properties we have investigated the AC conductivity in Yb$_2$CoMnO$_6$. The AC conductivity is calculated by using the relation $\sigma_{ac} = \epsilon_0 \omega \epsilon^{\prime\prime}$.\cite{sing, thakur} Fig. 14 shows the variation of AC conductivity with frequency i.e. $\sigma_{ac}$ vs $f$ at some selective temperatures in the range 50 K to 300 K. It is evident from the figure that at low frequencies the conductivity is independent of frequency and gives a plateau region at all temperatures. In this region of frequency, the conduction is mainly dominated by DC conductivity ($\sigma_{dc}$). However, at higher frequencies, the conductivity increases with increasing frequency. It is further to note that the plateau region marked by DC conductivity in Fig. 14 extends to the higher frequencies with increasing temperature. The frequency independent region also suggests that the hopping charges carrier are absent at low frequencies.

The variation of AC conductivity with absolute temperature i.e.  $\sigma_{ac}$ vs 10$^3$/T at some selective frequencies is shown in Fig. 15a. We observed that with increasing frequency the conductivity increases. To understand the conduction mechanism we have fitted the the conductivity with the following formula:\cite{kim, puli, chen}
\begin{eqnarray}
\sigma_{ac} = \sigma_{0}exp\left(\frac{-E_a}{k_BT}\right)
\end{eqnarray}
where $\sigma_{0}$ is pre-exponent factor, $k_B$ is Boltzmann constant and $E_a$ is the activation energy.
From the fitting parameters, we have calculated the activation energy. It is found that the activation energy E$_a$ increases with decreasing frequency as shown in Fig. 15b.

\section{Conclusion}
Yb$_2$CoMnO$_6$ nano-crystalline was successfully synthesis by sol-gel method. In the present study, we report structural, magnetic, dielectric and transport properties of nano-crystalline Yb$_2$CoMnO$_6$. The structural analysis shows that the sample is in single phase and adopt monoclinic crystal structure with \textit{P2$_1$/n} space group. Magnetization study reveals that the Yb$_2$CoMnO$_6$ is a ferromagnetic material that undergoes PM-FM phase transition around $T_c$ $\sim$56 K. The material also shows antiferromagnetic ordering at low temperature. Raman study shows spin-phonon coupling present in this material. Dielectric measurement shows a strong dispersion in dielectric constant and tangent loss shows a relaxation phenomenon. The impedance spectroscopy shows that Yb$_2$CoMnO$_6$ does not follow Debye's model. The Nyquist plot analysis shows non-Debye's behavior of Yb$_2$CoMnO$_6$. AC conductivity shows strong frequency dependence at the higher frequency limit. The conductivity analysis shows that the conduction mechanism involves quantum mechanical tunneling phenomenon.

\section{Acknowledgment}
We acknowledge MNIT Jaipur, India for XPS data and AIRF (JNU) for magnetic measurement facilities. We acknowledge UGC-DAE-Consortium Indore and Dr. V. G. Sathe for Raman data. We also acknowledge Dr. A. K. Pramanik for dielectric measurement and UPEA-II funding for LCR meter. Author Ilyas Noor Bhatti acknowledges University Grants Commission, India for financial support.


\begin{thebibliography}{99}
\bibitem{kfwang} K. F. Wang, J. M. Liu and Z. F. Ren, Adv. Phys. \textbf{58}, 321 (2009).
\bibitem{fiebig} M. Fiebig, T. Lottermoser, D. Meier and M. Trassin, Nat. Rev. Mater. \textbf{1}(8), 16046 (2016). 
 \bibitem{heron} J. T. Heron, M. Trassin, K. Ashraf, M. Gajek, Q. He, S. Y. Yang, D. E. Nikonov, Y-H. Chu, S. Salahuddin and R. Ramesh, Phys. Rev. Lett. \textbf{107}, 217202 (2011).
\bibitem{seidel} J. Seidel, L. W. Martin, Q. He, Q. Zhan, Y.-H. Chu, A. Rother, M. E. Hawkridge, P. Maksymovych, P. Yu, M. Gajek, N. Balke, S. V. Kalinin, S. Gemming, F. Wang, G. Catalan, J. F. Scott, N. A. Spaldin, J. Orenstein and R. Ramesh, Nat. Mater. \textbf{8}, 229 (2009).
\bibitem{hoff} T. Hoffmann, P. Thielen, P. Becker, L. Bohatý, and M. Fiebig, Phys. Rev. B \textbf{84}, 184404 (2011).
\bibitem{sal} E. K. H. Salje,  Chem. Phys. Chem. \textbf{11}(5) 940 (2010).
\bibitem{ravi} S. Ravi and C. Senthilkumar, Ceramics International \textbf{43}(16) 14441 (2017).
\bibitem{yong} F. Yong, Y. Shi-Ming, Q. Wen, W. Wei, W. Dun-Hui and D. You-Wei, Chin. Phys. B \textbf{23}(11), 117501 (2014).
\bibitem{lezai} M. Lezai and N. A. Spaldin, Phys. Rev. B \textbf{83}, 024410 (2011).
\bibitem{blasco} J. Blasco, J. L. García-Muñoz, J. García, J. Stankiewicz, G. Subías, C. Ritter and J. A. Rodríguez-Velamazán, Appl. Phys. Lett. \textbf{107}, 012902 (2015).
\bibitem{vilar} S. Yáñez-Vilar, E. D. Mun, V. S. Zapf, B. G. Ueland, J. S. Gardner, J. D. Thompson, J. Singleton, M. Sánchez-Andújar, J. Mira, N. Biskup, M. A. Señarís-Rodríguez and C. D. Batista,
Phys. Rev. B \textbf{84}, 134427 (2011).
\bibitem{banerjee} A. Banerjee, J. Sannigrahi, S. Giri, and S. Majumdar, Phys. Rev. B \textbf{98}, 104414 (2018).
\bibitem{blasco1} J. Blasco, J. Garcıa, G. Subıas, J. Stankiewicz, J. A. Rodrıguez-Velamazan,
C. Ritter, J. L. Garcıa-Munoz and F. Fauth, Phys. Rev. B \textbf{93}, 214401 (2016).
\bibitem{ilyas1} Ilyas Noor Bhatti, Imtiaz Noor Bhatti, R. N. Mahato and M. A. H. Ahsan, Physics Letter A \textbf{383}, 2326 (2019).
\bibitem{bhatti1} Imtiaz Noor Bhatti, R. S. Dhaka and A. K. Pramanik, Phys. Rev. B \textbf{96}, 144433 (2017).
\bibitem{bhatti2} Imtiaz Noor Bhatti, R. Rawat, A. Banerjee and A.K. Pramanik, J. Phys.: Condens. Matter \textbf{27}, 016005 (2014).
\bibitem{wang} X. Wang, W. Li, X. Wang, J. Zhang, L. Sun, C. Gao, J. Shang, Y. Hu and Q. Zhu, RSC Adv. \textbf{7}, 50753 (2017).
\bibitem{qiu} B. Qiu, W. Guo, Z. Liang, W. Xia, S. Gao, Q. Wang, X. Yu, R. Zhao and R. Zou, RSC Adv. \textbf{7},13340 (2017).
\bibitem{xia} H. Xia, D. Zhu, Z. Luo, Y. Yu, X. Shi, G. Yuan and J. Xie, Scientific Reports \textbf{3} 2978 (2013).
\bibitem{ida} T. Hishida, K. Ohbayashi, and T. Saitoh, J. Appl. Phys. \textbf{113}, 043710 (2013).
\bibitem{sachoo} R. C. Sahoo, D. Paladhi and T. K. Nath, Journal of Magnetism and Magnetic Materials \textbf{436}, 77 (2017).
\bibitem{cao} Y. Cao, W. Li, K. Xu, Y. Zhang, T. Ji, R. Zou, J. Yang, Z. Qin and J. Hu, J. Mater. Chem. A \textbf{2}, 20723 (2014).
\bibitem{sarma} S. Ch. Sarma, U. Subbarao, Y. Khulbe, R. Jana and S. C. Peter, J. Mater. Chem. A \textbf{5},23369 (2017).
\bibitem{tiana} Z. Tiana, L. Zhengc, Z. Li, J. Li and J. Wanga, Journal of the European Ceramic Society \textbf{36}, 2813 (2016).
\bibitem{ilyas2} Ilyas Noor Bhatti, R. N. Mahato, Imtiaz Noor Bhatti, and M. A. H. Ahsan, Physica B: Condensed Matter \textbf{558}, 59 (2019).
\bibitem{ilyas3} Ilyas Noor Bhatti, Imtiaz Noor Bhatti, R. N. Mahato and M. A. H. Ahsan, Ceramics International \textbf{46}, 46 (2020).
\bibitem{renu} Renu Gupta, Imtiaz Noor Bhatti abd A. K. Pramanik, J. Phys.: Condens. Matter \textbf{32}, 035803 (2020).
\bibitem{bhatti3} Imtiaz Noor Bhatti abd A. K. Pramanik, Physics Letters A \textbf{383}, 1806 (2019).
\bibitem{ilyas4} Ilyas Noor Bhatti, Rabindra Nath Mahato, Imtiaz Noor Bhatti and M.A.H. Ahsan, Materials Today: Proceedings, \textbf{17}, Part 1, 216 (2019).
\bibitem{bhar1} S. C. Bhargava, S. Singh and S.K. Malik, Journal of Magnetism and Magnetic Materials \textbf{311}, 594 (2007).
\bibitem{bhar2} S. C. Bhargava, S. Singh and S.K. Malik, Phys. Rev. B \textbf{71}, 104419 (2005).
\bibitem{chikazumi} S. Chikazumi, \textit{Physics of ferromagnetism. English edition prepared with the assistance of C.D. Graham, Jr (2nd ed.).} Oxford: Oxford University Press (2009)
\bibitem{cullity} B. D. Cullity and C. D. Graham,  \textit{Introduction to Magnetic Materials} John Wiley $\&$ Sons (2011)
\bibitem{ilive} M. N. Iliev, M. V. Abrashev, A. P. Litvinchuk, V. G. Hadjiev, H. Guo, and A. Gupta, Phys. Rev. B \textbf{75}, 104118 (2007).
\bibitem{meyer} C. Meyer, S. Hühn, M. Jungbauer, S. Merten, B. Damaschke, K. Samwer, and V. Moshnyaga, J. Raman Spectrosc. \textbf{48}, 46 (2017).
\bibitem{granado} E. Granado, A. García, J. A. Sanjurjo, C. Rettori, I. Torriani, F. Prado, R. D. Sanchez, A. Caneiro and S. B. Oseroff, Phys. Rev. B \textbf{60}, 11879 (1999).
\bibitem{laver} J. Laverdiere, S. Jandl, A. A. Mukhin, V. Yu. Ivanov, V. G. Ivanov, and M. N. Iliev, Phys. Rev. B \textbf{73}, 214301 (2006).
\bibitem{sch} A. Schonhals and F. Kremer, {\it Broadband Dielectric Spectroscopy} Springer-Verlag Berlin Heidelberg (2003)
\bibitem{kao} K. C. Kao, {\it Dielectric Phenomena in Solids: With Emphasis on Physical Concepts of Electronic Processes} Elsevier (2004).
\bibitem{mansuri} Amantulla Mansuri, Ilyas Noor Bhatti, Imtiaz Noor Bhatti and Ashutosh Mishra, Journal of Advanced Dielectrics  \textbf{08}, No. 04, 1850024 (2018).
\bibitem{kim} J. E. Kim, S. J. Kim, H. W. Choi and Y. S. Yang, J. Korean Phys. Soc. \textbf{42}, 1224 (2003).
\bibitem{elli} S. R. Elliott, Adv. Phys., \textbf{36}, 135 (1987).
\bibitem{bhatn} V. K. Bhatnagar and K. L.Bhatia, J. Non-Cryst. Solids, \textbf{119}, 212 (1990).
\bibitem{ali} H. A. M. Ali and M. A. Ibrahim, Materials Science-Poland, \textbf{34}(2), 386 (2016).
\bibitem{sing} A. Sing, A. Gupta and R. Chatterjee, Appl. Phys. Lett. \textbf{93}, 022902 (2008).
\bibitem{thakur} S. Thakur, R. Rai, Igor Bdikin and M.A. Valente, Ceramics Mater. Res. \textbf{19}, 1 (2016).
\bibitem{puli} V. S. Puli, C. Orozco, R. Picchini and C. V. Ramana, Materials Chemistry and Physics \textbf{184}, 82 (2016).
\bibitem{chen} C. Chen, P. Jost, H. Volker, M. Kaminski, M. Wirtssohn, U. Engelmann, K. Kruger, F. Schlich,
C. Schlockermann, R. P. S. M. Lobo and M. Wuttig, Phys. Rev. B \textbf{95}, 094111 (2017).
\end{thebibliography}
\end{document}